 \documentclass[final,3p,times,twocolumn]{elsarticle}
 
\usepackage{amssymb}
\usepackage{ulem}
\usepackage{color}

\begin{document}

\begin{frontmatter}

\title{$J/\psi$ production in $ep$ collisions at HERA}

\author{Mathias Butensch\"on}
%\ead{mathias.butenschoen@desy.de}
\address{{II.} Institut f\"ur Theoretische Physik, Universit\"at Hamburg,
Luruper Chaussee 149, 22761 Hamburg, Germany}

\begin{abstract}
This report is a short overview of experimental and theoretical results in $J/\psi$ production in electron-proton collisions at DESY HERA with special focus lying on recent developments in inelastic photoproduction.
\end{abstract}


\end{frontmatter}

\section{Introduction}
\label{Introduction}

Collisions of electrons or positrons with protons ($ep$) are among the particle reactions which have been intensively used to study different mechanisms for heavy quarkonium production. In this report we limit ourselves to the production of $J/\psi$ mesons, which have an experimentally very clean signature due to the large branching ratios of their leptonic decay modes. Let us start with some definitions. In $ep$ collisions the electron interacts with the proton almost entirely via a bremsstrahlung photon $\gamma$. In most reactions, this photon is quasi-real, its squared momentum $-Q^2$ being very small. This kinematic region is called {\it photoproduction}, while the region in which the photon retains a high virtuality $Q^2$ is called {\it leptoproduction}. A second way to classify the reactions is to look at the inelasticity variable $z=(p_{J/\psi}\cdot p_p)/(p_\gamma\cdot p_p)$. In the proton rest frame, $z$ is the ratio of the photon momentum taken over by the $J/\psi$ meson. At values $z\approx 1$, the $J/\psi$ absorbs nearly all of the photon momentum, and a rapidity gap can be observed. This kinematic range is called {\it elastic} or {\it diffractive} production. We call it {\it elastic} in case of a really exclusive $ep\to ep+J/\psi$ process, and {\it diffractive}, if during the reaction the initial proton evolves into a low lying nucleon resonance. The kinematic region with $z\lessapprox0.95$ is called {\it inelastic} production, and inelastic leptoproduction is called {\it deep inelastic scattering} (DIS).

\section{Inelastic $J/\psi$ production}

\begin{figure*}
\includegraphics[width=5.2cm]{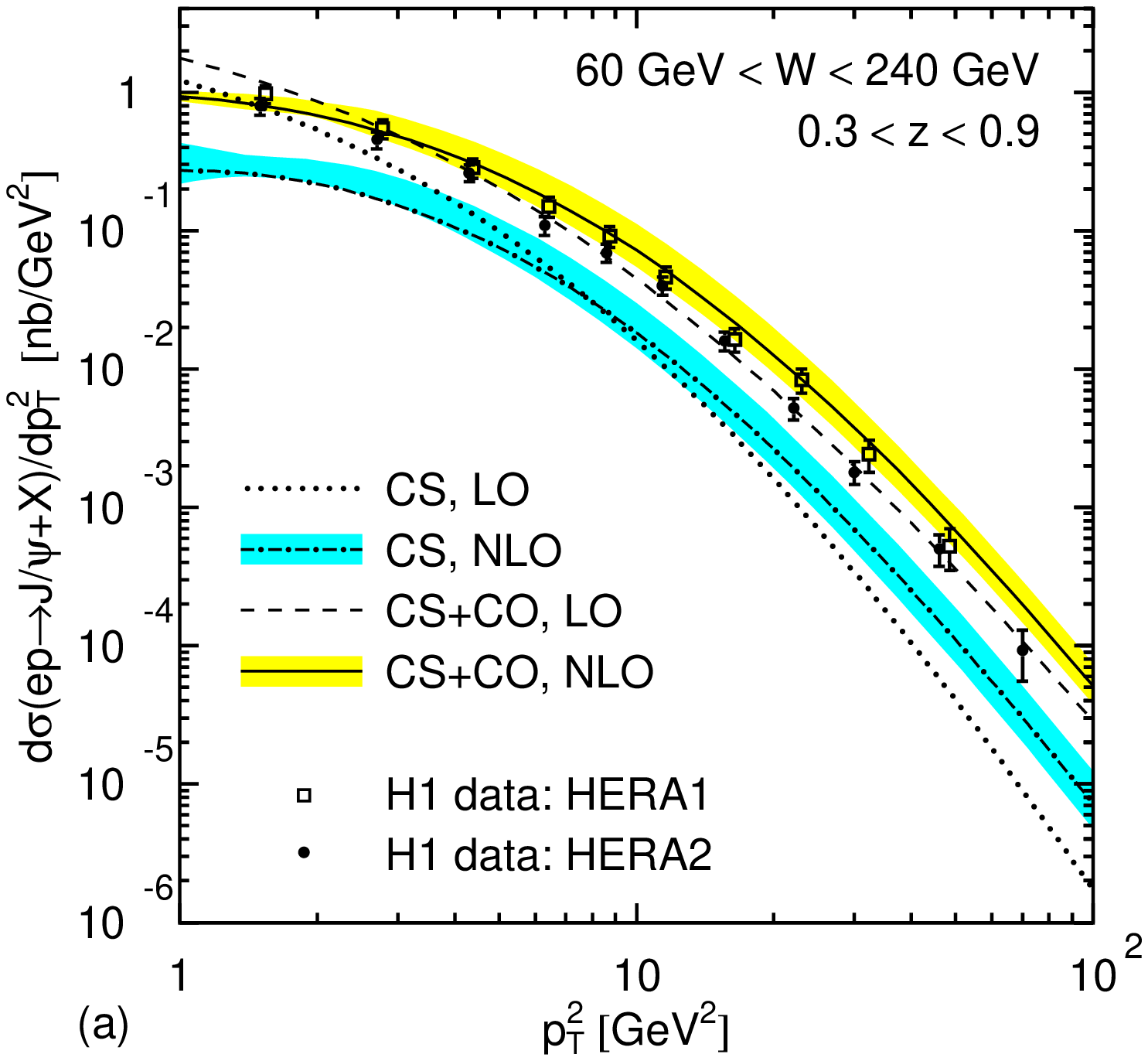} \hfill
\includegraphics[width=5.2cm]{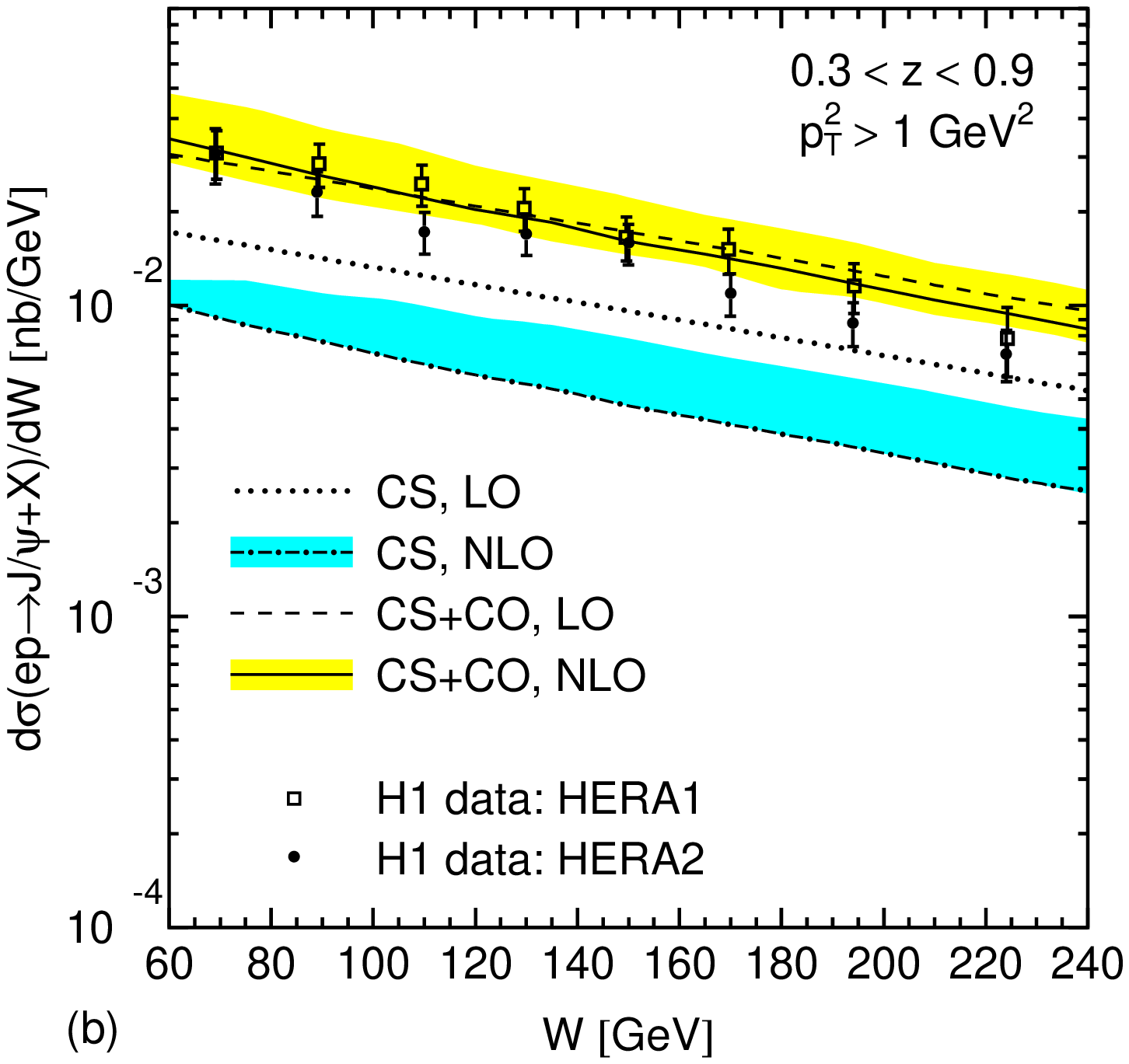} \hfill
\includegraphics[width=5.2cm]{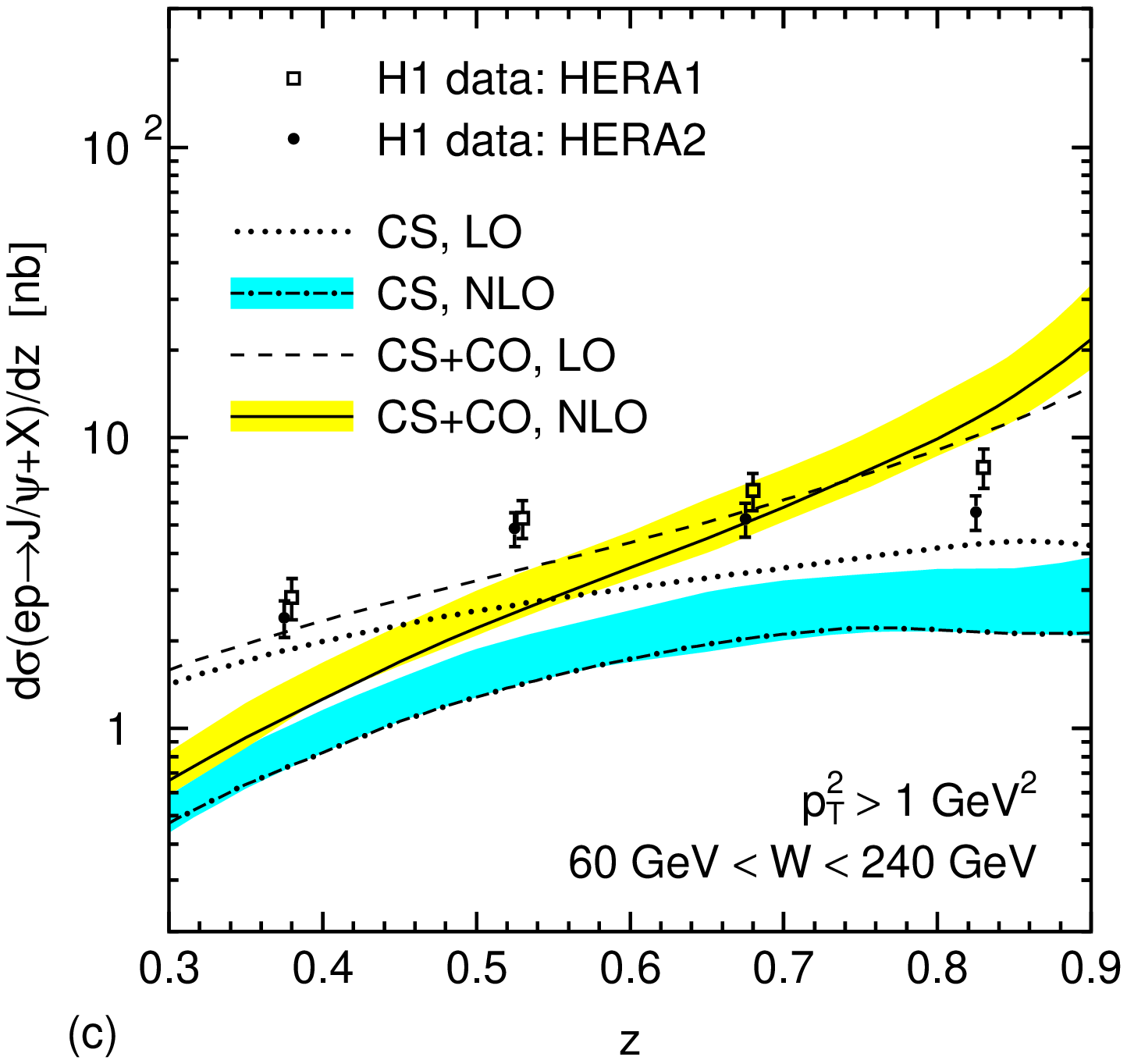}
\caption{\label{FigA} The transverse momentum squared $p_T^2$, photon-proton invariant mass $W$ and $z$ distributions of the inclusive inelastic $J/\psi$ photoproduction cross section. Leading order (LO) and next-to-leading order (NLO) predictions in NRQCD \cite{Butenschoen:2010rq} including and not including the color octet (CO) contributions are compared to H1 HERA1 \cite{Adloff:2002ex} and HERA2 \cite{Aaron:2010gz} data. The color octet LDMEs used were obtained by a common fit to the $p_T$ distributions of CDF Tevatron data \cite{Acosta:2004yw} and the data of plot (a). These plots are taken over from \cite{Butenschoen:2010rq}.}
\end{figure*}

There are different models on the market which aim to describe the inelastic production of heavy quarkonia. In all these models, the photon interacts with only one parton inside the proton. In the usual {\it collinear} factorization, the initial parton is assumed to be on shell and collinear to the proton. We obtain the hadronic cross section by folding the partonic ones with parton distribution functions (PDFs) $f_{i/p}(x)$, where $x$ is the longitudinal momentum fraction of the parton $i$, which can be a gluon or (anti-)quark. Within the framework of {\it nonrelativistic QCD} (NRQCD) \cite{Bodwin:1994jh}, the partonic $J/\psi$ production cross section factorizes further according to
\begin{eqnarray}
\lefteqn{d\sigma(ep\to J/\psi+X) = \sum_{i,n} \int dx \,f_{i/p}(x)} \nonumber \\
&\times& d\sigma(ei\to c\overline{c}[n]+X) \; \langle{\cal O}^{J/\psi}[n] \rangle
\end{eqnarray}
into perturbative short distance cross sections $d\sigma(ei\to c\overline{c}[n]+X)$ and long distance matrix elements (LDMEs) $\langle{\cal O}^{J/\psi}[n]\rangle$, usually fitted to experimental data. $n$ can be any intermediate Fock state, including color octet (CO) states. According to NRQCD, each $\langle{\cal O}^{J/\psi}[n]\rangle$ scales with a definite power of the relative quark velocity $v$, which then serves as an additional expansion parameter besides $\alpha_s$. The main contribution stems from the $n={^3S}_1^{[1]}$ term, which equals the 
{\it color singlet model} (CSM) prediction and the next-to-leading contributions stem from intermediate $^1S_0^{[8]}$, $^3S_1^{[8]}$, and $^3P_J^{[8]}$ CO states.

The idea behind using the {\it $k_T$ factorization} \cite{Gribov:1984tu} method is that in quarkonium production processes with typical hard scattering scales much lower than the collision energies, the longitudinal momentum fraction $x$ is so small that the parton's transverse momentum $k_T$ should not be neglected. Therefore the initial parton is off shell. The partonic cross section, which is so far only evaluated at leading order in $\alpha_s$, is then convoluted with {\it unintegrated}, $k_T$ dependent PDFs, which are derived from the usual gluon PDFs either in a BFKL \cite{Kuraev:1977fs} or DGLAP \cite{Gribov:1972ri} approach. Usually, only color singlet (CS) contributions are considered for $J/\psi$ production calculations in this formalism. The Monte Carlo program CASCADE \cite{Jung:2000hk} simulates initial gluon radiation in such an approach.

\subsection{Inelastic photoproduction yield}

Born level calculations for the $J/\psi$ photoproduction yield have been performed in the color singlet model \cite{Berger:1980ni} already in 1980, and the calculation including the color octet states \cite{Cacciari:1996dg} has been performed soon after the invention of NRQCD. While these leading order predictions do include {\it resolved} contributions, in which the photon interacts via its hadronic content, next-to-leading order (NLO) contributions are so far only known for the {\it direct} processes.

Figure~\ref{FigA} shows the result of our recent NLO NRQCD analysis \cite{Butenschoen:2010rq} of the inclusive $J/\psi$ photoproduction cross section using values for the CO LDMEs obtained from a combined NLO fit to the $p_T$ distributions of CDF Tevatron \cite{Acosta:2004yw} and H1 HERA \cite{Adloff:2002ex,Aaron:2010gz} $J/\psi$ production data. The CS contributions alone undershoot the H1 data \cite{Adloff:2002ex,Aaron:2010gz} typically by a factor of three, which is however a much lesser deviation than in the hadroproduction case. Our CS results are in agreement with previous NLO CSM calculations \cite{Kramer:1994zi,Artoisenet:2009xh,Chang:2009uj}. Apparent differences to the early work of \cite{Kramer:1994zi} are only due to different parameter choices. The sum of the CS and CO contributions on the other hand describes the data sufficiently well. As for the $z$ distribution, the cross section at low $z$ is expected to rise once we include the resolved photon contribution. Near the high $z$ endpoint region, the NRQCD expansion is understood to break down, and the NRQCD series could be resummed via the introduction of universal {\it shape functions} \cite{Beneke:1997qw}, possibly in the context of soft collinear effective theory \cite{Fleming:2006cd}. The main achievement of our recent work \cite{Butenschoen:2010rq}, which is a continuation of \cite{Butenschoen:2009zy}, is to show that $J/\psi$ photoproduction at H1 HERA and hadroproduction at CDF Tevatron, PHENIX RHIC and CMS LHC can consistently be described by a unique set of CO LDMEs.

\begin{figure}
\centering
\includegraphics[width=5cm]{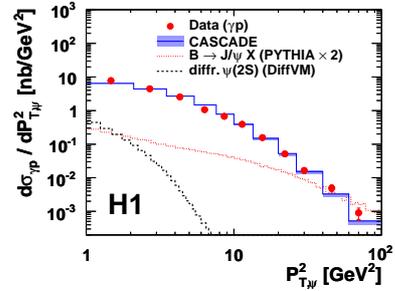}
\caption{\label{FigB} The H1 HERA2 data \cite{Aaron:2010gz} compared to CASCADE \cite{Jung:2000hk} predictions. Relative to figure \ref{FigA}a, the cross section is divided by an averaged photon flux factor 0.102. This plot is taken from \cite{Aaron:2010gz}.}
\end{figure}

Figure~\ref{FigB} shows the Monte Carlo results of CASCADE compared to the same H1 HERA2 data \cite{Aaron:2010gz}. It gives very good results, even though, according to the author, there is no kind of tuning of CASCADE parameters to the data involved. The $k_T$ factorization calculation \cite{Lipatov:2002tc} also shows good agreement with H1 \cite{Adloff:2002ex} and ZEUS \cite{Chekanov:2002at} HERA1 data.

\subsection{Polarization in inelastic photoproduction}

\begin{figure*}
\begin{minipage}[c]{5.2cm}
\includegraphics[width=5.2cm]{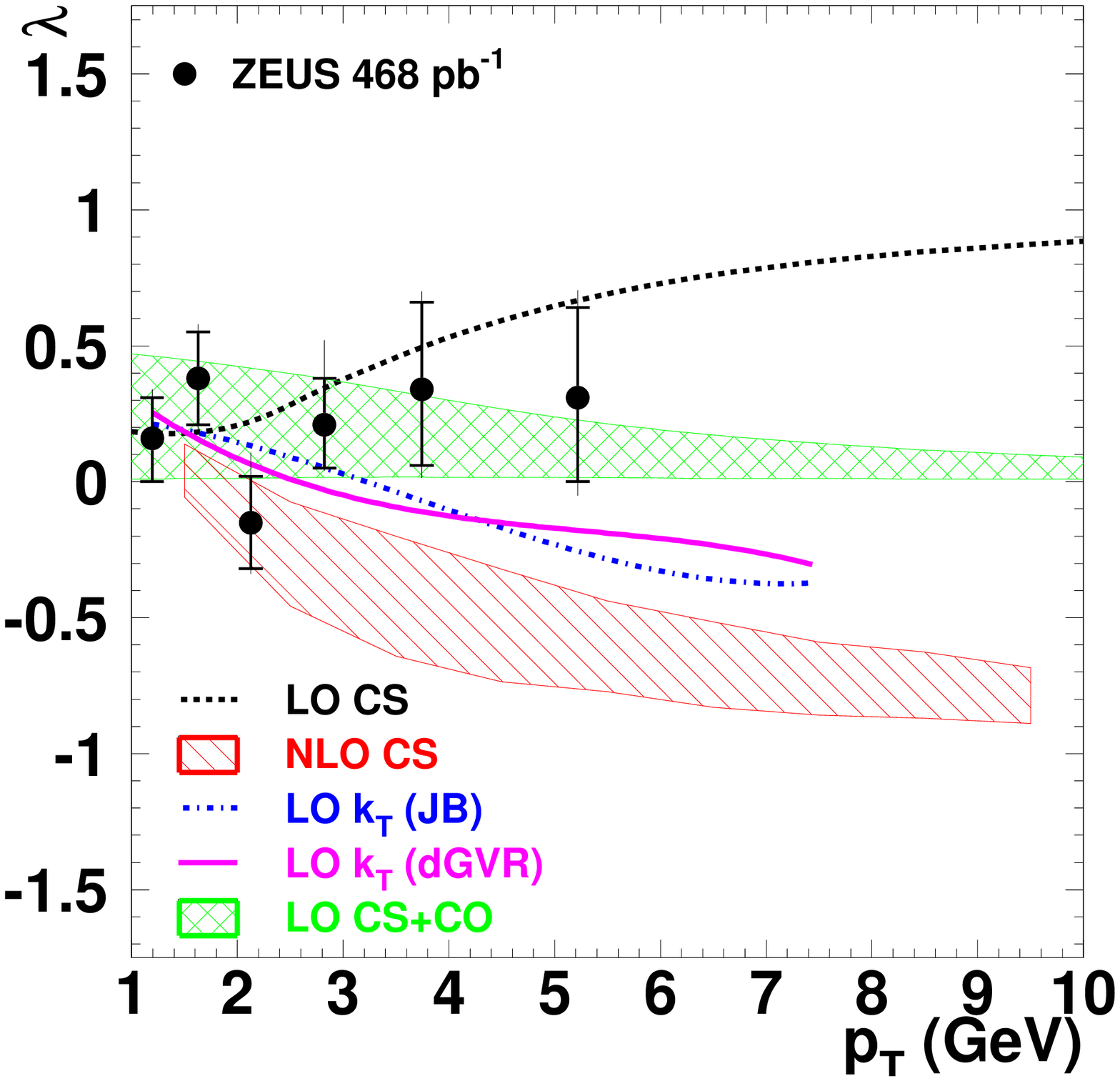}
\end{minipage}\hfill
\begin{minipage}[c]{5.2cm}
\includegraphics[width=5.2cm]{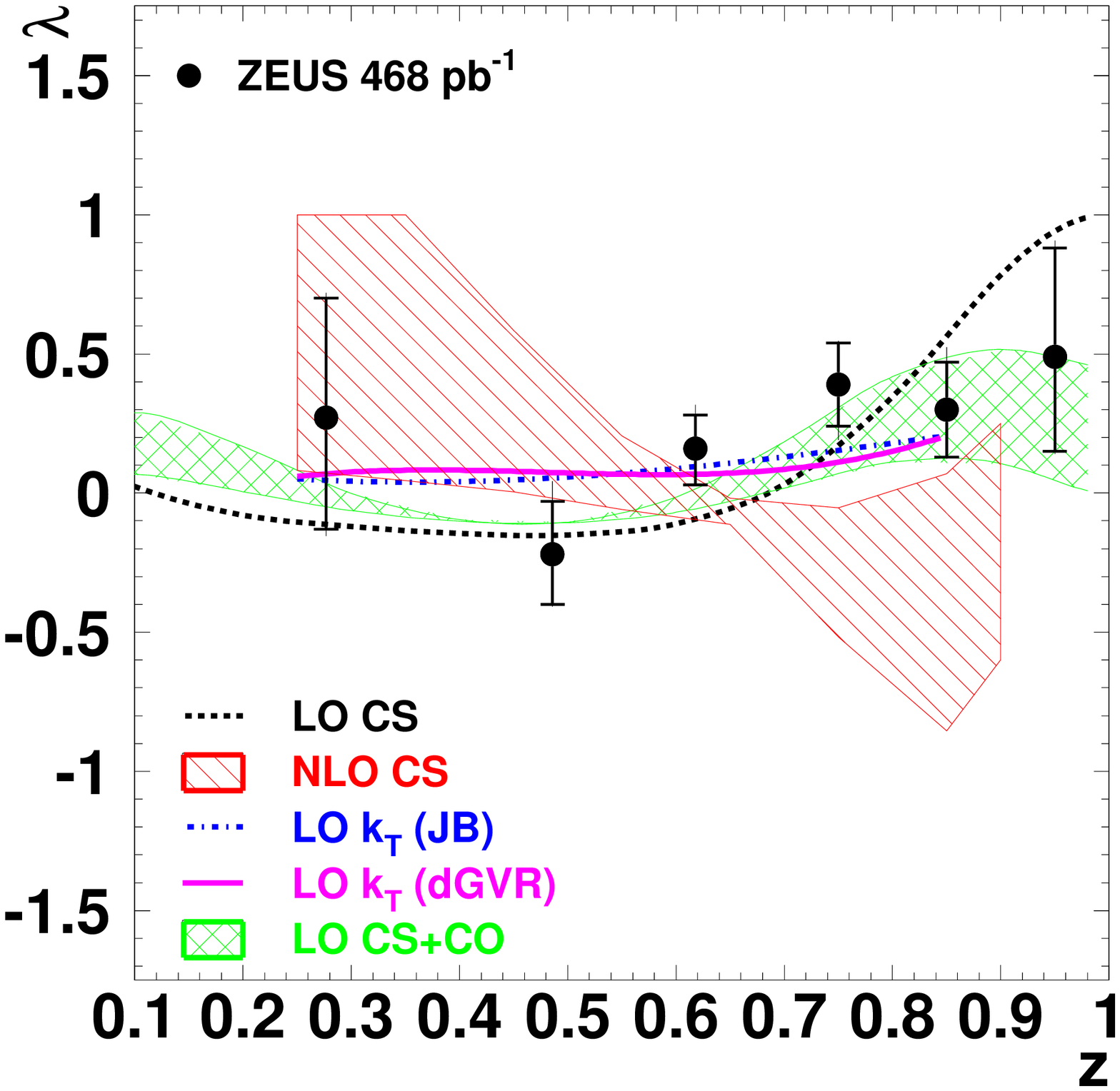}
\end{minipage}\hfill
\begin{minipage}[c]{5.2cm}\vspace{-12pt}
\caption{\label{FigC} The helicity parameter $\lambda$ in the target frame in inelastic $J/\psi$ photoproduction as functions of the transverse $J/\psi$ momentum $p_T$ and the inelasticity $z$. The kinematical ranges are 50~GeV $<W<$ 180~GeV, $0.4<z<1$ for the $p_T$ distribution and $p_T>1$~GeV for the $z$ distribution. The ZEUS data \cite{:2009br}, taken 1996-2007, are compared to leading order (LO) and next-to-leading order (NLO) CSM predictions \cite{Artoisenet:2009xh}, an LO prediction including CO states \cite{Beneke:1998re}, as well as predictions with $k_T$ factorization \cite{Baranov:2008zzc} using different unintegrated gluon PDFs. Only the LO CS and LO CS+CO curves include resolved processes. These plots are taken from \cite{:2009br}.}
\end{minipage}
\end{figure*}

\begin{figure*}
\begin{minipage}[c]{5.2cm}
\includegraphics[width=5.2cm]{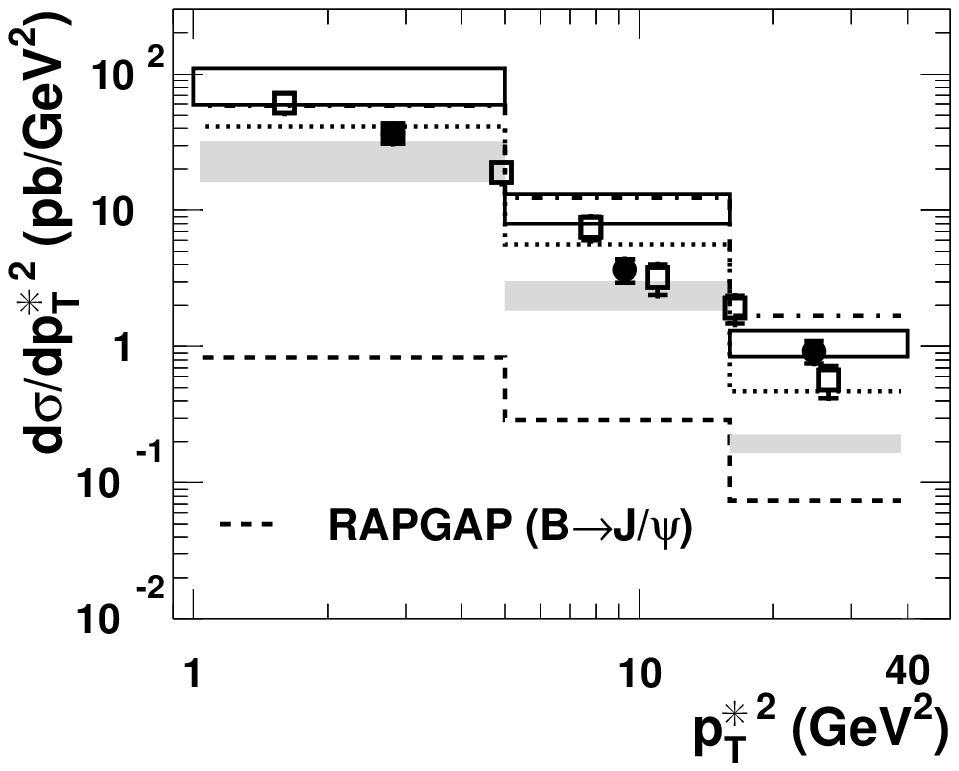}
\end{minipage}\hfill
\begin{minipage}[c]{5.2cm}
\includegraphics[width=5.2cm]{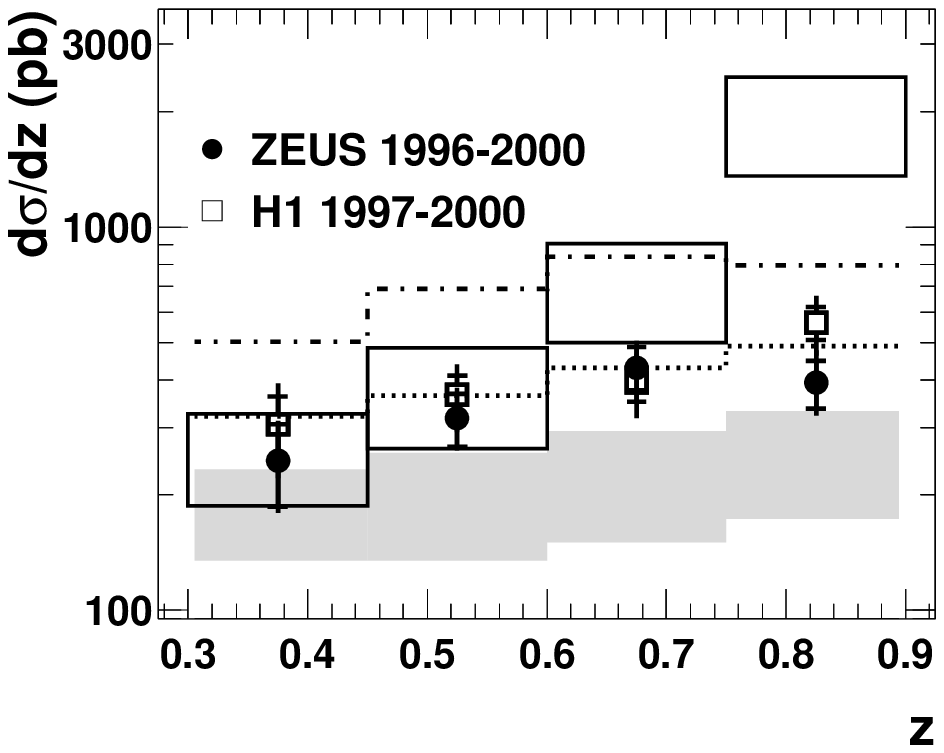}
\end{minipage}\hfill
\begin{minipage}[c]{5.2cm}\vspace{-14pt}
\caption{\label{FigD} The differential inclusive deep inelastic $J/\psi$ production cross section with respect to the $J/\psi$ transverse momentum squared in the photon-proton rest frame $p_T^{\ast 2}$ and $z$. The kinematical ranges are $2$~GeV$^2<Q^2<100$~GeV$^2$, $50$~GeV $<W<225$~GeV, $0.3<z<0.9$ and $p_T^{\ast 2}>1$~GeV$^2$. The ZEUS data \cite{Chekanov:2005cf} and the H1 data \cite{Adloff:2002ey} are compared with leading order CS (gray bands) and CS+CO (open bands) predictions \cite{Kniehl:2001tk}, $k_T$ factorization prediction \cite{Lipatov:2002tc} (dotted lines) and CASCADE \cite{Jung:2000hk} results (dash-dotted lines). These plots are taken from \cite{Chekanov:2005cf}.}
\end{minipage}
\end{figure*}

As for polarization observables, theoretical predictions have been made for the color singlet contributions at NLO \cite{Artoisenet:2009xh,Chang:2009uj}. The color octet contributions, however, are so far only known at leading order \cite{Beneke:1998re}. In figure~\ref{FigC}, these are compared to recent ZEUS data \cite{:2009br} together with the $k_T$ factorization prediction \cite{Baranov:2008zzc}. Here, the polarization parameter $\lambda$ is defined according to
\begin{equation}
 \frac{d\sigma}{d\cos\theta} = 1 + \lambda \cos^2 \theta,
\end{equation}
where $\theta$ is the polar angle of the decay-muons in the target reference frame. Values $\lambda=+1$ (-1) correspond to fully transversely (longitudinally) polarized $J/\psi$ mesons. Like in the H1 polarization analysis in \cite{Aaron:2010gz}, which is done in the helicity and Collins-Soper frame, both theoretical and experimental errors are still too large to draw definite conclusions about the agreement for the different production mechanisms. Specifically, no NLO calculation including CO contributions has been performed yet.

\subsection{Deep inelastic scattering}

In contrast to inelastic photoproduction, DIS squared matrix elements contain a further scale $Q^2$, which may be one reason, why until now only leading order calculations have been performed for $J/\psi$ production in DIS. In figure~\ref{FigD}, the results of the NRQCD calculation \cite{Kniehl:2001tk}, the $k_T$ factorization calculation \cite{Lipatov:2002tc} and CASCADE \cite{Jung:2000hk} predictions are compared to ZEUS \cite{Chekanov:2005cf} and H1 data \cite{Adloff:2002ey}. Due to the large theoretical uncertainties, conclusions are difficult to be drawn, but the CS prediction in collinear factorizations again seems to lie rather below the data, while we again see an overshoot of the CO contributions at high $z$.

As for $J/\psi$ polarization in DIS, no measurement and only the CSM calculation \cite{Yuan:2000cn} have been done so far.

\begin{figure*}
\begin{minipage}[c]{5.2cm}
\includegraphics[width=5.2cm]{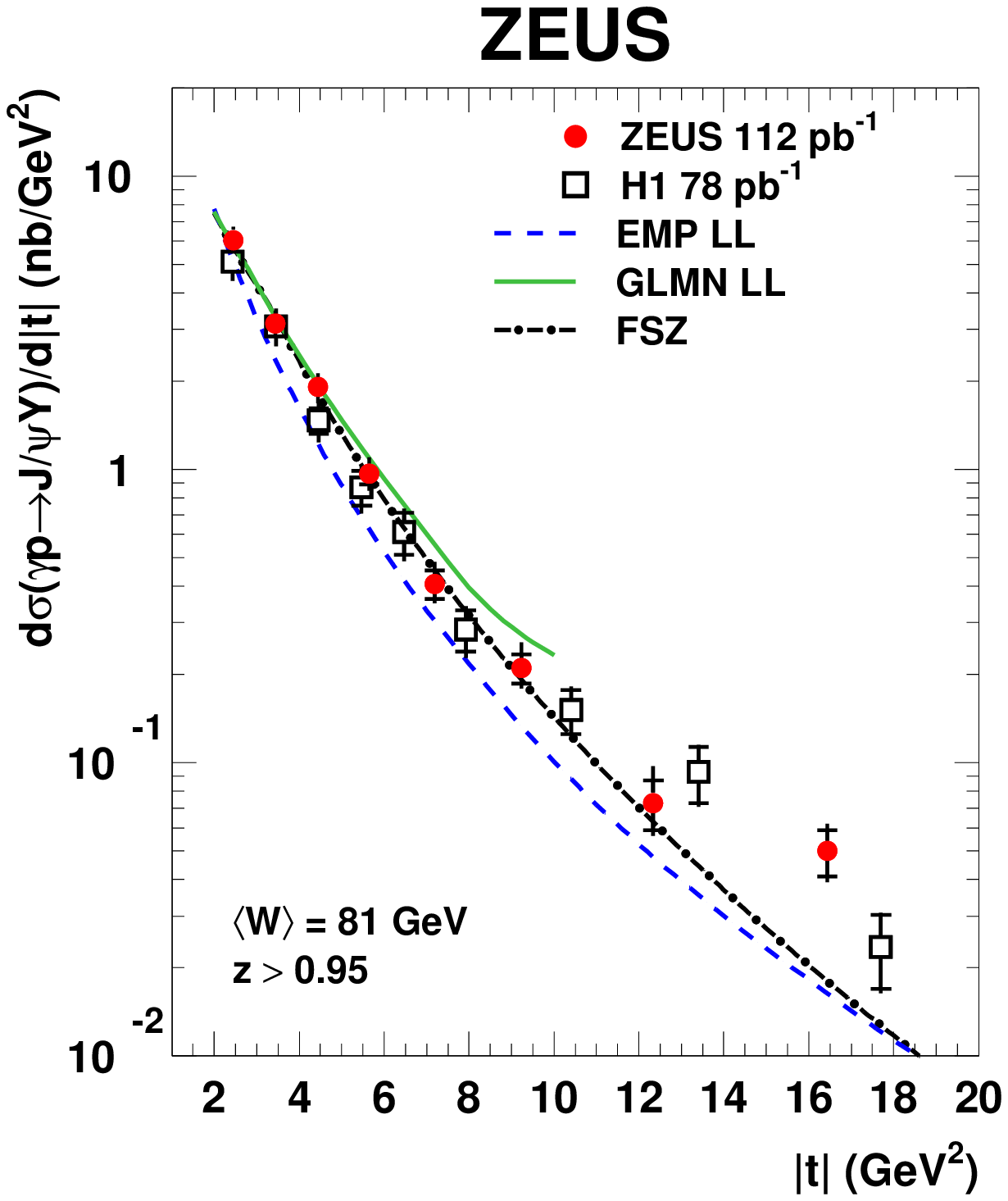}
\end{minipage}\hfill
\begin{minipage}[c]{5.2cm}
\includegraphics[width=5.2cm]{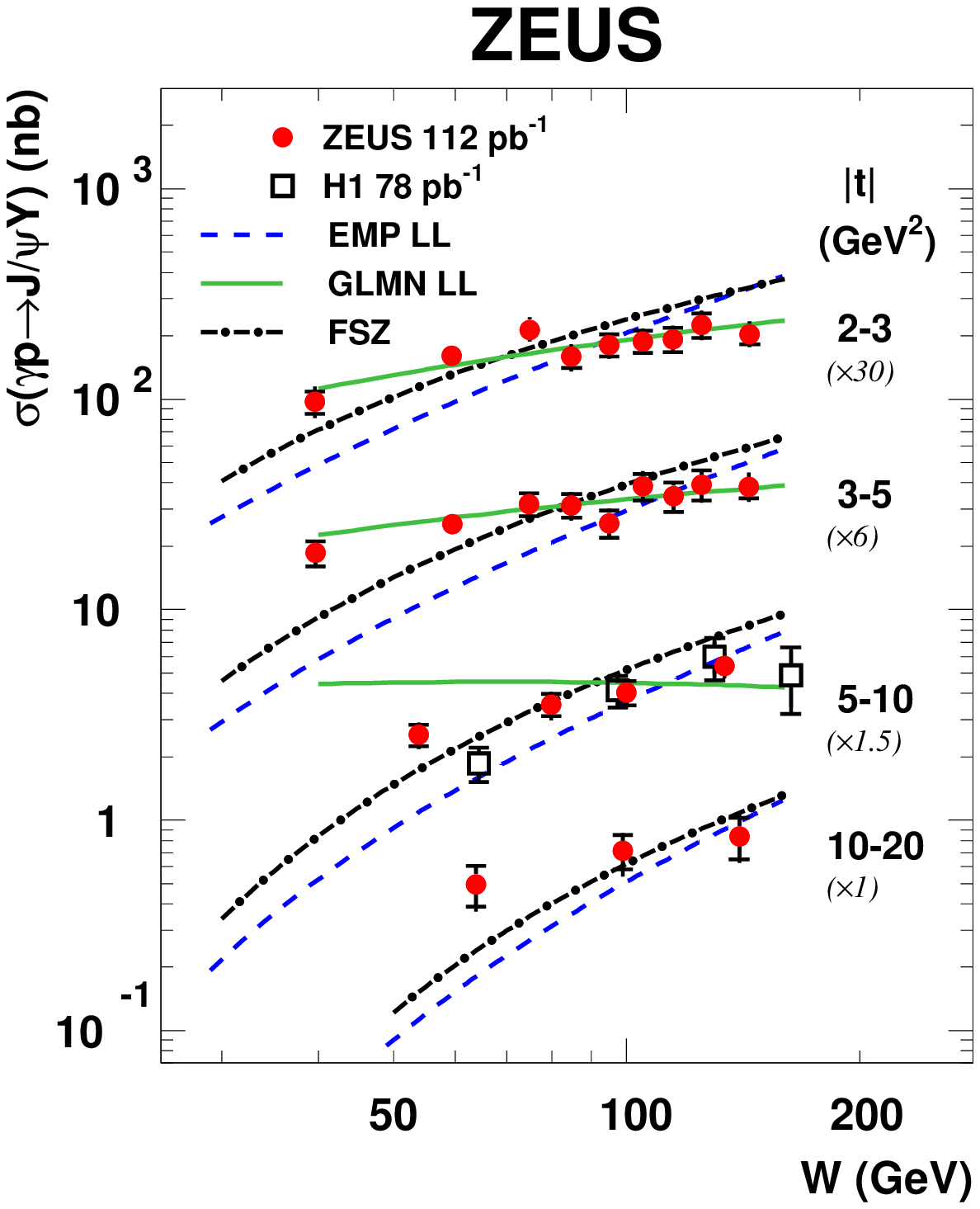}
\end{minipage}\hfill
\begin{minipage}[c]{5.2cm}\vspace{-12pt}
\caption{\label{FigE} The differential diffractive $J/\psi$ photoproduction cross section with respect to the absolute value of the squared momentum transfer at the proton vertex $|t|$ and the photon-proton invariant mass $W$. The ZEUS \cite{:2009tu} and H1 \cite{Aktas:2003zi} data, both taken 1996-2000, are compared to theoretical predictions \cite{Enberg:2002zy} (EMP), \cite{Gotsman:2001ne} (GLMN) and \cite{Frankfurt:2008er} (FSZ). These plots are taken from \cite{:2009tu}.}
\end{minipage}
\end{figure*}

\section{Elastic and diffractive $J/\psi$ production}

In elastic or diffractive $J/\psi$ production, the photon couples to the $c\overline{c}$ pair, which then in turn interacts with the proton through the exchange of an object with vacuum quantum numbers, for example two or more gluons. Within the two-gluon-ladder model, the leading logarithms of this ladder can be resummed using BFKL \cite{Kuraev:1977fs} evolution.

In figure~\ref{FigE}, diffractive $J/\psi$ photoproduction cross sections measured at ZEUS \cite{:2009tu} and H1 \cite{Aktas:2003zi} are compared to theoretical predictions using a DGLAP \cite{Gotsman:2001ne} or BFKL \cite{Enberg:2002zy,Frankfurt:2008er} approach. Measurements for elastic $J/\psi$ production have been performed by the ZEUS \cite{Chekanov:2002xi,Chekanov:2004mw} and H1 \cite{Adloff:2000vm,Aktas:2005xu} collaborations for both photoproduction \cite{Adloff:2000vm,Chekanov:2002xi,Aktas:2005xu} and leptoproduction \cite{Chekanov:2004mw,Aktas:2005xu}.

\section{Conclusions} 

Collisions of electrons or positrons and protons at HERA have been a very important testing field for different models for the production of heavy quarkonia, of $J/\psi$ in particular. The results must, of course, be seen in combination with quarkonium production in other particle reactions, for example in hadron collisions. The simple collinear factorization color singlet model seems to be disfavored by inelastic $J/\psi$ production data. Therefore, promising attempts have been made to explain the data either within NRQCD via the introduction of intermediate color octet states or within the $k_T$ factorization approach. However, none of the models is ruled out yet.

\end{document}